\documentclass[conference,letterpaper]{IEEEtran}

\IEEEoverridecommandlockouts
\usepackage{cite}
\usepackage{svg}
\usepackage{amsmath,amssymb,amsfonts}
\usepackage{algorithmic}
\usepackage{graphicx}
\usepackage{textcomp}
\usepackage{xcolor}
\usepackage{colortbl}
\usepackage[normalem]{ulem}

\usepackage{xspace} 

\usepackage{tabu}
\usepackage{dcolumn} 
\usepackage{booktabs}
\usepackage{enumitem}
\usepackage{multirow}
\usepackage{wrapfig}

\usepackage{graphicx}
\usepackage{tikz}
\usepackage{pgfplots}

\usepackage{etoolbox}

\usepackage[ruled]{algorithm2e}

\pgfplotsset{compat=1.17}
\usepackage{hyperref}
\usepackage[nameinlink,capitalise,noabbrev]{cleveref}
\usepackage{scalerel}
\usepackage{rotating}
\usepackage{soul}
\usepackage{comment}
\usepackage{float}

\usetikzlibrary{arrows,bending,calc,positioning,shadows,shapes}

\definecolor{linkcolor}{rgb}{0.7,0.1,0.1}
\definecolor{citecolor}{rgb}{0,0.65,0}
\definecolor{urlcolor}{rgb}{0,0,0.65}
\hypersetup{colorlinks=true, linkcolor=linkcolor, urlcolor=urlcolor, citecolor=citecolor}

\usepackage{array}

\newcommand{\parhead}[1]{\vspace{5pt plus 2pt minus 2pt}\par\noindent\textbf{#1}\hspace{.4em plus .2em minus .2em}} 

\definecolor{darkviolet}{HTML}{9400D3}
\definecolor{darkgreen}{rgb}{0,0.65,0}

\newcommand{\al}{\textsf{AssemblyLine}\xspace}
\newcommand{\instr}[1]{\texttt{#1}\xspace}
\newcommand{\code}[1]{\texttt{#1}\xspace}
\newcommand{\ots}{off-the-shelf\xspace}

\newcommand{\cryptopt}{{Crypt\-Opt}\xspace}
\newcommand{\fiat}{{Fiat Cryptography}\xspace}
\newcommand{\fiatir}{{Fiat~IR}\xspace}
\newcommand{\rls}{{RLS}\xspace}

\newcommand{\approachRRR}{{R3-validation}\xspace}
\newcommand{\entry}{Optimizer\xspace}

\newcommand{\assembly}{assembly\xspace}
\newcommand{\assemblies}{assembly codes\xspace}
\newcommand{\xassembly}{x86-64~\assembly}
\newcommand{\xassemblies}{x86-64~\assemblies}

\newcommand{\json}{\texttt{JSON}\xspace}

\newcommand{\of}{\texttt{OF}}

\newcommand{\gccVersion}{11.3.0\xspace}
\newcommand{\clangVersion}{14.0.0\xspace}

\SetCommentSty{mycommfont}

\newcommand{\docolor}[2]{{\setlength{\fboxsep}{0pt}%
  \ifthenelse{\equal{#1}{A}}{\colorbox{green!15}{#2}}{%
  \ifthenelse{\equal{#1}{B}}{\colorbox{blue!15}{#2}}{%
  \ifthenelse{\equal{#1}{C}}{\colorbox{yellow!15}{#2}}{%
  \ifthenelse{\equal{#1}{D}}{\colorbox{red!15}{#2}}{%
    #1#2%
  }}}}%
}}%
{
\catcode`\!\active

}

\usepackage[utf8]{inputenc}
\usepackage{newunicodechar}

\newunicodechar{λ}{$\lambda$}
\newunicodechar{∀}{$\forall$}
\newunicodechar{⌊}{$\lfloor$}
\newunicodechar{⌋}{$\rfloor$}

\begin{document}

\title{\cryptopt: Automatic Optimization\\of Straightline Code\vspace{-0.8mm}}

\newcommand{\uoa}{University of Adelaide}
\newcommand{\gatech}{Georgia Institute of Technology}

\newcommand{\aff}[1]{$^#1$}
\author{
  Joel Kuepper\aff{1},
  David Wu\aff{1},
  Andres Erbsen\aff{2},
  Jason Gross\aff{2},
  Owen Conoly\aff{2},
  Chuyue Sun\aff{3},
  Samuel Tian\aff{2},\\
  Adam Chlipala\aff{2},
  Chitchanok Chuengsatiansup\aff{4},
  Daniel Genkin\aff{5},
  Markus Wagner\aff{6},
  Yuval Yarom\aff{1}\\\vspace{-1.2mm}
  \\
  \aff{1}\,University of Adelaide,\hspace{2em}\aff{2}\,Massachusetts Institute of Technology,\hspace{2em}\aff{3}\,Stanford University,\\\aff{4}\,University of Melbourne\hspace{5em}\aff{5}\,Georgia Tech,\hspace{5em}\aff{6}\,Monash University\vspace{-2.5mm}
}

\maketitle

\begin{abstract}

Manual engineering of high-performance implementations typically consumes many resources and requires in-depth knowledge of the hardware.
Compilers try to address these problems; however, they are limited by design in what they can do.
To address this, we present \cryptopt, an automatic optimizer for long stretches of straightline code.
Experimental results across eight hardware platforms show that \cryptopt achieves a speed-up factor of up to 2.56 over current \ots compilers.

\end{abstract}



\sloppy

\vspace{-2.5mm}
\section{Introduction}
\label{s:introduction}

Superscalar processors in recent years have become very complex, and inherent CPU properties make it hard to reason about the performance of a given assembly code~\cite{variabilityCPU,stabilizer,threadContextSwitch,producingWrongData,benchmarkPrecision,bokhari2019mind}.
As such,  optimizing code performance for a specific CPU microarchitecture is not trivial.
Typical compiler optimizations focus on control flow rather than long stretches of straightline code~\cite{AhoSU86}.
Moreover, they often utilize peephole optimizers~\cite{AhoSU86,BERGMANN2003141} with heuristics of replacement patterns to statically optimize small sections of code.

We present \cryptopt, an automatic optimizer for long stretches of straightline arithmetic code.
\cryptopt recasts compilation as a combinatorial optimization problem, with runtime as the cost function, and utilizes techniques from the field of search-based software
engineering~\cite{HARMAN2001833} to optimize.
We observe that rather simple techniques, 
such as a random local search (\rls)~\cite{doerr2019theory}, can 
produce faster code 
than current \ots compilers, such as GCC and Clang, even when used with their highest optimization settings.
Two observations are the key enablers for our approach. 
First, specializing in straightline code simplifies code analysis.
This simplicity, in turn, enables \cryptopt to explore many optimization options,
such as reordering operations, where conventional compilers, including  GCC and Clang, tend to be more conservative.
Second, by actually running the generated code on the target hardware we can optimize specifically for particular architectures, while treating the CPU  as a black box, removing
the need for complicated, error-prone, and lengthy modeling.

Cryptographic code typically follows the constant-time programming paradigm to mitigate timing side-channel attacks~\cite{CauligiSBJHJS17,MolnarPSW05,AlFardanP13}.
As such, it tends to contain long stretches of straightline arithmetic, which we use as our first use case: 
we use \cryptopt to generate high-performance cryptographic code,\footnote{The full version of this paper shows that this code is also formally verified~\cite{KuepperJ23arXiv}.} optimized for eight CPU architectures,
achieving speedups of up to 87\% across platforms and up to 156\% in single cases. 
We also show that we can optimize on a per-architecture level. 
That is, we can generate a solution on one platform (optimized for the same platform) that outperforms every other solution optimized on (and for) other platforms.

We believe that this technique can serve as a foundation for future engineering of high-performance implementations.
Rather than employing reverse engineering and processor modeling, we employ search algorithms and performance measurements. 
That is, we simply run our optimizer on future processors to generate optimized code with minimal effort.
\cryptopt is open-source, available at \url{https://0xADE1A1DE.github.io/CryptOpt}.

\section{Overview}

We now sketch \cryptopt's input language and outline how it works at a high level.
Then, we focus on how a user would use it for their own architecture or for their own input functions.
We conclude this section with a detailed description of the inner workings of \cryptopt.

\subsection{Input Language}
\label{s:inputlang}

\cryptopt reads the description of an input function in an intermediate language (see below). 
This input language is sufficient to describe even large expressions including modular arithmetic using bitwise operations.
It even allows to materialize $\phi$-nodes via a conditional-move ($\mathsf{cmovznz}$) operation.

\vspace{-1mm}
{\footnotesize
  $$\begin{array}{rrcl}
    \textrm{Variable} & x \\
    \textrm{Binary integer} & b \\
    \textrm{Operand} & e &::=& x \mid b \\
    \textrm{Operator} & o &::=& ! \mid \& \mid * \mid + \mid - \mid \; << \; \mid \; = \; \mid \; >> \; \mid \; \sim \; \mid \\
      &&&  \mathsf{or} \mid \mathsf{addcarryx} \mid \mathsf{cmovznz} \mid \mathsf{mulx} \mid \\
      &&&  \mathsf{static\_cast} \mid \mathsf{subborrowx} \\
    \textrm{Expression} & E &::=& \mathsf{return} \; e \mid x, \ldots, x \leftarrow o(e, \ldots, e); E
  \end{array}$$
}

\subsection{High-Level Concept}

\begin{figure}
    \centering
   
        \begin{tikzpicture}[-latex,very thick]

        \tikzstyle{box}     = [draw, inner sep=1em];
        \tikzstyle{og}      = [fill=blue!30];
        \tikzstyle{mut}     = [fill=orange!30];
        
        \node (in) [text width=9em] {Input function\\ (\json specification)};
        \node (ir)   [below =1em of in, box, og] {IR};
        \node (code) [below =1em of ir,      box, og] {\texttt{asm} code};
        \draw (in.south) -| (ir.north);
        \draw (ir.south) -| (code.north);

        \node (mutir) [right=10em of ir,     box, mut] {IR};
        \draw (ir.east) -- (mutir.west) node[midway, draw, fill=white, rounded rectangle ] (mutate){Mutate};

        \node (mutcode) [below=1em of mutir, box, mut] {\texttt{asm} code};
        \draw (mutir.south) -| (mutcode.north);

        \node (run) [below=5em of mutate, draw, rounded rectangle, inner sep=0.5em, text width=9em]{measure performance};
        \draw (mutcode.west) -- ([xshift=-2em]run.north east);
        \draw (code.east) -- ([xshift=2em]run.north west);

    \end{tikzpicture}
    \caption{\cryptopt high-level concept: Input specification is parsed into an IR and assembly code. The initial IR is then mutated (shown in orange). The performance of both
        candidates is then compared.}
    \label{f:high_lvl_concept}
\end{figure}
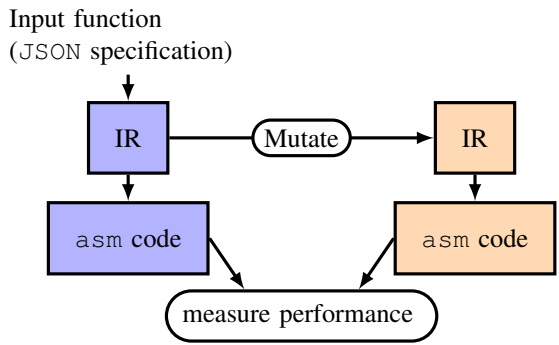

\cryptopt parses a function in the input language (specified as a \json file) into an internal representation (IR).
From this IR, \cryptopt derives a base implementation candidate in assembly  (blue in \cref{f:high_lvl_concept}).
At a high level, \cryptopt uses a \rls for optimizing the code.  \cref{f:high_lvl_concept} shows the basic step of the \rls algorithm.
Specifically, \cryptopt mutates the current IR, changing the instruction selected for implementing an IR operation or the order of operations to derive an alternative assembly implementation (orange in \cref{f:high_lvl_concept}).
Next, \cryptopt measures the performance of both assembly implementations, i.e.\ of the previous best and of the mutant, and  keeps the faster candidate.
This \emph{mutate-measure-select} step is repeated iteratively, resulting in an optimized implementation of the input function.
While measuring the performance, \cryptopt also checks the functional result of the candidates against a \texttt{C}-compiled ``ground truth.''

\subsection{User Workflow}\label{s:workflow}
The current application of \cryptopt is for optimizing  cryptographic code.
In particular, we focus on cryptographic primitives such as finite-field arithmetic, as widely used in elliptic-curve cryptography (e.g.\ in TLS and Bitcoin~\cite{libsecp256k1}) and postquantum cryptography (though schemes like SIKE~\cite{sike}).
To that end, we integrate with \fiat~\cite{ErbsenPGSC19}, which can generate both a \json specification of a field-arithmetic routine as well as a (decently optimized) \texttt{C} reference.
In this particular use case, \cryptopt can be invoked with a ``curve--method'' combination, e.g.\ using the command \texttt{./CryptOpt} \texttt{--bridge} \texttt{fiat} \texttt{--curve} \texttt{curve25519} \texttt{--method} \texttt{square}.
\fiat is then consulted internally to generate the required code.
To use \cryptopt for other functions, which are not produced by \fiat, the user should use the \texttt{--bridge manual} parameter and provide both a \json specification of the input function and a \texttt{C}
reference of the same function as the ground truth, e.g.\ \texttt{./CryptOpt --bridge manual --jsonFile ./example.json --cFile ./example.c}.
Note that \cryptopt is not a compiler plugin. Rather, \cryptopt is a self-contained tool to generate \xassembly from the input specification outlined in \cref{s:inputlang}.

\begin{figure}[t]
    \centering
    \input{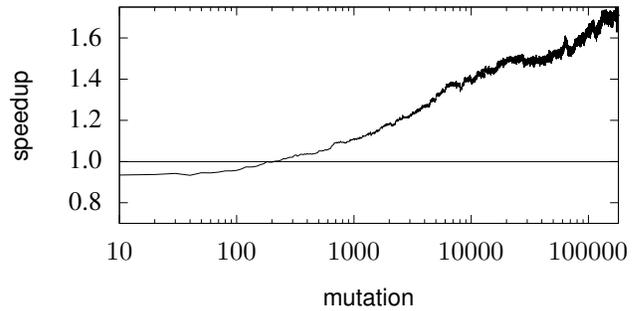}
    \caption{Optimization progress of SIKEp434--square on Intel Core i7-10710U, showing the relative performance gain over Clang as a function of the number of tested mutations.}
    \label{f:conv}
\end{figure}

While optimizing, the user receives regular  status updates on the console, providing information on, e.g.\ the number of instructions used to implement the function in the current version, the number of memory spills, or the relative speedup
compared to the \texttt{C}-compiled ground truth.
Further, we also generate a PDF file showing how the optimization progresses over time.
See \cref{f:conv} for an example.

\subsection{Detailed Internals}
\label{s:details}

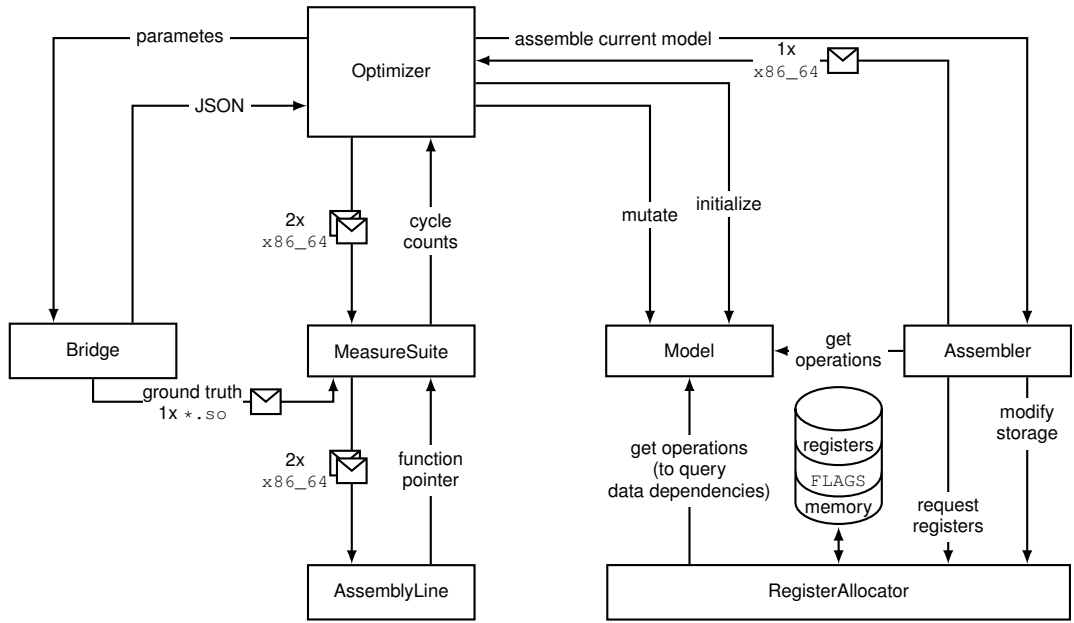
\begin{figure*}[t]
\centering
\begin{tikzpicture}[
    font=\sffamily,
    -latex,
    thick,
    yscale=.6,xscale=.7,
    database/.style n args={3}{scale=7,cylinder,aspect=0.5,draw,rotate=90,path picture={
            \draw[-] (path picture bounding box.160) to[out=180,in=180] (path picture bounding box.20);
            \draw[-] (path picture bounding box.200) to[out=180,in=180] (path picture bounding box.340);
            \node[] (c) {#2};
            \node[above=.3em of c ] (a) {#1};
            \node[below=.1em of c ] (b) {#3};
        }
    },
    envelope/.style={rectangle, fill=white, draw, scale=2, inner xsep=.4em, path picture={
            \draw[-,draw] (path picture bounding box.north west) -- (path picture bounding box.center) -- (path picture bounding box.north east){};
    }},
    xenvelope/.style n args={2}{inner xsep=2.8em, scale=3, path picture={
            \node[envelope] (env) {};
            \node[inner xsep=0, left=2pt of env.west, anchor=east,align=center,fill=white] (text) {#1\\#2};
    }}
    ]

    \scriptsize
    \tikzstyle{class}     = [draw=black, inner sep=1em, minimum width=9em];



    \matrix[column sep=7em, row sep=10em] (m)  at (9,7){
                                & \node[class, minimum height=7em] (opt) {Optimizer};                                                                          \\
    \node[class] (br) {Bridge}; & \node[class] (ms)                      {MeasureSuite};      & \node[class] (model)  {Model}; & \node[class] (as) {Assembler};\\
                                & \node[class] (al)                      {AssemblyLine};                                                                       \\
};
    \coordinate  (modelAl) at ($(model)!0.5!(as)$);
    \node[class, minimum width=25em] (ra)      at (al -| modelAl)          {RegisterAllocator};

    \node[database={registers}{\texttt{FLAGS}}{memory}, above=4.5em of ra, anchor=center] (db) {};
    \draw[latex-latex] (ra) to ($(db.west)+(0,.1)$);
    \draw[-latex] let \p1=(ra.north), \p2=($(as.south)    + (0.75,0)$) in (\p2) -- (\x2, \y1)  node [near start, align=center, fill=white]{modify\\storage};
    \draw[-latex] let \p1=(ra.north), \p2=($(as.south)    - (0.75,0)$) in (\p2) -- (\x2, \y1)  node [near end,   align=center, fill=white]{request\\registers};
    \draw[latex-] let \p1=(ra.north), \p2=($(model.south)           $) in (\p2) -- (\x2, \y1)  node [midway,     align=center, fill=white]{get operations\\(to query\\data dependencies)};
    \draw[-latex] (as.west)  -- (model.east)  node [midway,     align=center, fill=white]{get\\operations};

    \draw[latex-] let \p1=(al.north), \p2=($(ms.south)    + (0.75,0)$) in (\p2) -- (\x2, \y1)  node [midway,     align=center, fill=white]{function\\pointer};
    \draw[-latex] let \p1=(al.north), \p2=($(ms.south)    - (0.75,0)$) in (\p2) -- (\x2, \y1)  node [midway,     ] (xma) {};
    \node[envelope] at ($(xma.north west)  + (.01,.01) $){};
    \node[xenvelope={2x}{\texttt{x86\_64}}] at (xma){};

    \draw[latex-] let \p1=(ms.north), \p2=($(opt.south)   + (0.75,0)$) in (\p2) -- (\x2, \y1)  node [midway,     align=center, fill=white]{cycle\\counts};
    \draw[-latex] let \p1=(ms.north), \p2=($(opt.south)   - (0.75,0)$) in (\p2) -- (\x2, \y1)  node [midway,     ] (xom) {};
    \node[envelope] at ($(xom.north west)  + (.01,.01) $){};
    \node[xenvelope={2x}{\texttt{x86\_64}}] at (xom){};

    \draw[latex-] ($(br.north) - (0.75,0)$)|- ($(opt.west)   + (0,0.75)$) node [near end,     align=center, fill=white]{parametes};
    \draw[-latex] ($(br.north) + (0.75,0)$)|- ($(opt.west)   - (0,0.75)$) node [near end,     align=center, fill=white]{JSON};

    \draw[-latex] (br.south)               |- ++(0,-.5)  -| ($(ms.south west)   + (0.5,0)$) node [near start,   ](xbm){} ;
    \node[xenvelope={ground truth}{1x \texttt{*.so}}] at ($(xbm) + (1,0)$){};

    \draw[latex-] ($(model.north) - (0.75,0)$)|- ($(opt.east)   - (0,0.75)$) node [near start,     align=center, fill=white]{mutate};
    \draw[latex-] ($(model.north) + (0.75,0)$)|- ($(opt.east)   - (0,0.25)$) node [near start,     align=center, fill=white]{initialize};

    \draw[latex-] ($(as.north) + (0.75,0)$)|- ($(opt.east)   + (0,0.75)$) node [very near end,     align=center, fill=white]{assemble current model};
    \draw[-latex] ($(as.north) - (0.75,0)$)|- ($(opt.east)   + (0,0.25)$) node [midway,  ](xoa){};
    
    \node[xenvelope={1x}{\texttt{x86\_64}}] at ($(xoa) - (2,0)$){};

\end{tikzpicture}
\caption{Component diagram of \cryptopt.\label{f:inner}}
\end{figure*}

We now describe the inner workings of \cryptopt, shown in \cref{f:inner}.
The user invokes \cryptopt with the parameters described in \cref{s:workflow}.
Upon invocation, the provided parameters are parsed, and the control is handed to \entry.
\entry selects the required Bridge---a module which generates both the required function description in a \json format and a shared object
(\texttt{*.so}-file) from the \texttt{C} reference, which is used for correctness checks while measuring.
\entry then initializes Model, a component used as the single point of truth for the IR.
The initialization procedure of Model also preprocesses the operations specified in the input description.
This includes
\begin{itemize}[nosep,leftmargin=*]
\item \textbf{Instruction scheduling}: analyzing the data flow and subsequently generating an initial ordering of the operations.
\item \textbf{Instruction selection}: assigning a template to each operation specifying which \xassembly instruction(s) implement it.
\end{itemize}
After that, \entry invokes Assembler to assemble the current IR.
Assembler initializes Register Allocator, which is the component maintaining the  virtual state of the CPU \texttt{FLAGS} register, general-purpose registers, and stack memory.
The Register Allocator is initialized with caller-save registers holding (unknown) live values and calling-convention-based registers, holding function parameters.
Assembler processes the operations from Model according to the current order,
producing the \xassembly according to the assigned template.
While doing that, Assembler consults Register Allocator to get empty registers or modify the storage locations of intermediate values.
Register Allocator will, in case it needs to spill a variable to memory, get the next operations from Model to determine which value is most suitable to spill.
Assembler eventually returns the \xassembly string to \entry, which stores it temporarily.
\entry then invokes the mutate function on Model.
This will (randomly) change the IR in one of two ways:
change the order in which the operations are implemented,
or  change the template assigned to one of the operations.
\entry will then invoke Assembler again to generate the assembly code of the mutated IR.
\entry then uses  MeasureSuite to compare the performance of the implementation of the original IR with that of the mutated IR.
MeasureSuite uses \al~\cite{assemblyline} to assemble two \xassemblies and perform a version of the \approachRRR~\cite{Bokhari2020}, as outlined in \cref{s:measure}.
Finally, MeasureSuite returns the  measurement results to \entry, which compares the results and decides whether to accept the mutated IR as the new base implementation.
Otherwise, \entry discards the mutation and proceeds to test another one.
After a predefined number of mutations has been tried,  \cryptopt writes the last base implementation to disk.

\section{Measurement}\label{s:measure}

Measuring precise execution times of short stretches of code is challenging because it is affected by a large
number of noise factors~\cite{variabilityCPU,threadContextSwitch,producingWrongData,benchmarkPrecision,bokhari2019mind}.
Bokhari et al.~\cite{Bokhari2020} compared validation approaches in the context of energy-consumption optimization.
We adopt their \approachRRR approach with two modifications.
First, we do not restart the computer after each evaluation, because we do not observe any measurement drift over time.
Second, we use a random scheduling of program variants  instead of following a strict order of measurements to reduce effects of learned execution orders by the CPU.

Recall that we need to assess which of the two candidate \xassembly implementations is faster.
Our  measurement procedure randomly selects one candidate and measures it in a tight loop.
Then, we measure the check function (ground truth) in a tight loop.
The number of iterations in each of those loops depends on the measured code: Slow stretches are repeated fewer times than fast stretches.
With this dynamic, we can skew the nominal number of cycles measured into a range where the performance counter gives enough granularity.
Empirically, aiming for roughly 10\,000 cycles provides a good trade-off in terms of sensitivity because it amplifies the execution time differences enough to be detected, yet is robust enough not to be misled
by the environmental factors.

For stability, we repeat this procedure multiple times, each with randomly selected  input values and compare the median of these experiments to determine which of the candidates to keep.
We empirically find that 31 repetitions provide sufficiently stable results on our systems.
The user can adjust the number to suit the running environment, e.g.\ a higher number may be required for noisy systems.

\begin{table}
    \scriptsize
    \begin{center}
        \caption{Overview of target machines used in the experiments}
        \label{tab:machines}
        \begin{tabular}{p{2.25cm}@{\hspace{.25cm}}p{1.48cm}@{\hspace{.25cm}}p{4.15cm}}
            \toprule
             CPU & $\mu$-arch & Mainboard \\
            \midrule
             AMD Ryzen Theadripper 1900X  & Zen 1          & ASUS ROG STRIX X399-E Gaming \\
             AMD Ryzen~7 5800X               & Zen 3          & Gigabyte B550 AORUS ELITE V2 \\
             AMD Ryzen~9 5950X               & Zen 3          & Gigabyte X570 GAMING X \\
             Intel Core i7-6770HQ            & Skylake-H      & Intel NUC6i7KYB \\
             Intel Core i7-10710U            & Comet Lake-U   & Intel NUC10i7FNB \\
             Intel Core i9-10900K            & Comet Lake-S   & Gigabyte H470 HD3 \\
             Intel Core i7-11700KF           & Rocket Lake-S  & ASRock Z590 Pro4 \\
             Intel Core i9-12900KF           & Alder Lake-S   & Micro-Star PRO Z690-A Wifi (MS-7D25) \\
            \bottomrule
        \end{tabular}
    \end{center}
    \vspace{-1em}
\end{table}

\section{\cryptopt in the real world}
\parhead{Evaluation setup.}
We evaluate \cryptopt on eight different platforms, summarized in \cref{tab:machines}.
On each machine, we generate \xassembly code with \cryptopt for the multiply and square methods of nine different prime fields from \fiat.

\parhead{Optimization process.}
Optimization takes between 36 and 70 wall-clock hours to generate those 18 primitives, depending on the machine.
The length of the produced code depends on the compiled primitive and varies from fewer than 100 instructions to almost 1000 instructions.
The average length of the best implementations generated is shown in \cref{tab:avgInstrCount}.

\begin{table}[htb]
        \centering
        \footnotesize
        \caption{Average instruction count, rounded to nearest integer}
        \label{tab:avgInstrCount}
        \begin{tabular}{lrr}
            \toprule
            Primitive & Multiply & Square \\
            \midrule 
            Curve25519       & 170       & 121   \\
            NIST P-224       & 221       & 219   \\
            NIST P-256       & 204       & 200   \\
            NIST P-384       & 710       & 698   \\
            SIKEp434         & 986       & 965   \\
            Curve448         & 588       & 405   \\
            NIST P-521       & 542       & 338   \\
            Poly1305         &  76       &  61   \\
            secp256k1        & 233       & 224   \\
            \bottomrule
        \end{tabular}
        \vspace{-.5em}
\end{table}

\parhead{Optimized code performance.}
\cref{tab:res-avg} shows the geometric mean over those eight platforms of the speedup of  \cryptopt-generated code vs.\ machine code generated by
\ots compilers Clang version \clangVersion and GCC version \gccVersion at the highest optimization settings (\code{-O3 -mtune=native -march=native}).

\begin{table}[t]
	\small
	\begin{center}
	\caption{Geometric means of \cryptopt vs.\ off-the-shelf compilers. 
  }\vspace{-1mm}
	\label{tab:res-avg}
    \begin{tabular}{lcccccc@{}}

		\toprule
			      & \multicolumn{2}{c}{Multiply} & & \multicolumn{2}{c}{Square}\\
			                \cmidrule{2-3}                  \cmidrule{5-6} 
			Curve & Clang & GCC & & Clang & GCC \\
		\midrule
			Curve25519  & \cellcolor{     blue!100}\color{white}{$1.19$} & \cellcolor{     blue!100}\color{white}{$1.14$} &  & \cellcolor{     blue!100}\color{white}{$1.14$} & \cellcolor{     blue!100}\color{white}{$1.18$}&\\
			P-224       & \cellcolor{     blue!100}\color{white}{$1.31$} & \cellcolor{     blue!100}\color{white}{$1.87$} &  & \cellcolor{     blue!100}\color{white}{$1.24$} & \cellcolor{     blue!100}\color{white}{$1.84$}&\\
			P-256       & \cellcolor{     blue!100}\color{white}{$1.27$} & \cellcolor{     blue!100}\color{white}{$1.79$} &  & \cellcolor{     blue!100}\color{white}{$1.30$} & \cellcolor{     blue!100}\color{white}{$1.85$}&\\
			P-384       & \cellcolor{     blue!100}\color{white}{$1.12$} & \cellcolor{     blue!100}\color{white}{$1.66$} &  & \cellcolor{     blue!76 }\color{white}{$1.08$} & \cellcolor{     blue!100}\color{white}{$1.60$}&\\
			SIKEp434    & \cellcolor{     blue!100}\color{white}{$1.30$} & \cellcolor{     blue!100}\color{white}{$1.70$} &  & \cellcolor{     blue!100}\color{white}{$1.29$} & \cellcolor{     blue!100}\color{white}{$1.83$}&\\
			Curve448    & \cellcolor{     blue!22 }\color{black}{$1.02$} & \cellcolor{   orange!52 }\color{black}{$0.95$} &  & \cellcolor{   orange!4  }\color{black}{$1.00$} & \cellcolor{   orange!8  }\color{black}{$0.99$}&\\
			P-521       & \cellcolor{     blue!100}\color{white}{$1.20$} & \cellcolor{     blue!58 }\color{white}{$1.06$} &  & \cellcolor{     blue!100}\color{white}{$1.25$} & \cellcolor{     blue!95 }\color{white}{$1.11$}&\\
			Poly1305    & \cellcolor{     blue!94 }\color{white}{$1.10$} & \cellcolor{     blue!100}\color{white}{$1.15$} &  & \cellcolor{     blue!79 }\color{white}{$1.09$} & \cellcolor{     blue!100}\color{white}{$1.16$}&\\
			secp256k1   & \cellcolor{     blue!100}\color{white}{$1.34$} & \cellcolor{     blue!100}\color{white}{$1.73$} &  & \cellcolor{     blue!100}\color{white}{$1.32$} & \cellcolor{     blue!100}\color{white}{$1.74$}&\\
		\bottomrule
		\end{tabular}\vspace{-2mm}
	\end{center}
\end{table}

\parhead{Side-channel resistance.}
Code generated by \fiat is timing-side-channel-secure.
\cryptopt will only optimize with different instruction scheduling, instruction selection, and register allocation.
\cryptopt will not change algorithmic structures\footnote{We do optimize with simple forms of strength reduction such as replacing multiplication by constants with a combination of
additions and bit-shifts.}.
We implement \fiatir's \instr{cmovznz}-operation with Intel's dedicated \instr{cmovCC} instruction.
As such, code based on \fiat optimized with \cryptopt is inherently timing-side-channel-secure.

\section{Related Work}
Next, we describe other work in the broader realm of automatically optimizing code.
We start with superoptimization, which targets the smallest pieces of code, and turn our focus then onto the field of peephole optimization, which considers larger chunks of instructions.

\parhead{Superoptimization.}
In 1987, Henry Massalin coined the term ``superoptimizer'' to describe his tool for exhaustive enumeration of all possible programs to implement a given function~\cite{Massalin87}.
Because exhaustive enumeration can require a huge computational effort, 
the key idea making this feasible is the use of a probabilistic test set, which rejects the majority of incorrect candidates.
His superoptimizer was able to generate programs of 12 instructions after several hours of running (on a 16MHz 68020 computer).
Since then, superoptimizers have evolved significantly: Souper~\cite{SasnauskasCCKTR17} can synthesize new optimizations on the LLVM IR, but as such they cannot exploit target-specific properties.
Denali~\cite{JoshiNR02} uses solvers to generate provably shortest programs, but it can only be applied to rather short program sequences in the range of tens of instructions.
STOKE~\cite{Schkufza0A13} and its extensions~\cite{Schkufza0A14, SharmaSCA13, SharmaSCA15} can synthesize new programs from scratch and optimize them, only focusing on very small kernels of loops.

\parhead{Peephole optimization.}
Instead of synthesizing new and optimal very short programs, 
peephole optimizers use a sliding window on instructions (the peephole) and replace sets of existing instructions with more performant alternatives~\cite{AhoSU86, COOPER2012597,BERGMANN2003141}.
The replacement is usually done based on a predefined rule set (applying only to short instruction sequences), which itself is based on heuristics for estimating which set of
instructions is likely to be more performant or shorter than an alternative.

One possible next step can be to find those heuristics automatically~\cite{Bansal06, PekhimenkoB10} and then to apply this new knowledge to the target code.
However, those rules are still applied statically, i.e. without taking the actual effects on runtime into account,
and they are typically applied over the entire code.

With \cryptopt, we overcome those limitations by first only applying a mutation locally and second by measuring the (side) effects of every mutation.

\section{Summary and Outlook}
We present \cryptopt, a tool for generating optimized assembly code by combining simple techniques.
As present, \cryptopt tackles distinctive characteristics of straightline cryptographic code, achieving a significant improvement over mainstream compilers.
In the future, we expect that these techniques can be generalized to other domains of compilation.
Moreover, it would be interesting to test if replacing \rls with more advanced optimization strategies would improve \cryptopt's run time and results.

\section*{Acknowledgments}
  This research was supported by 
  the Air Force Office of Scientific Research (AFOSR) under award number FA9550-20-1-0425; 
  the ARC DECRA DE200101577; 
  the ARC Discovery Projects DP200102364 DP210102670; 
  the Blavatnik ICRC at Tel-Aviv University; 
  CSIRO's Data61;
  the National Science Foundation under grants CNS-1954712 and CNS-2130671; 
  the National Science Foundation Expedition on the Science of Deep Specification award~CCF-1521584; 
  and gifts from
  Amazon Web Services,
  AMD,
  Cisco,
  Facebook,
  Google, 
  Intel, and the
  Tezos Foundation.

\IEEEtriggeratref{16}
\bibliographystyle{IEEEtranS}
\bstctlcite{IEEEexample:BSTcontrol}
\bibliography{cryptopt}

\begin{thebibliography}{10}
\providecommand{\url}[1]{#1}
\csname url@samestyle\endcsname
\providecommand{\newblock}{\relax}
\providecommand{\bibinfo}[2]{#2}
\providecommand{\BIBentrySTDinterwordspacing}{\spaceskip=0pt\relax}
\providecommand{\BIBentryALTinterwordstretchfactor}{4}
\providecommand{\BIBentryALTinterwordspacing}{\spaceskip=\fontdimen2\font plus
\BIBentryALTinterwordstretchfactor\fontdimen3\font minus
  \fontdimen4\font\relax}
\providecommand{\BIBforeignlanguage}[2]{{%
\expandafter\ifx\csname l@#1\endcsname\relax
\typeout{** WARNING: IEEEtranS.bst: No hyphenation pattern has been}%
\typeout{** loaded for the language `#1'. Using the pattern for}%
\typeout{** the default language instead.}%
\else
\language=\csname l@#1\endcsname
\fi
#2}}
\providecommand{\BIBdecl}{\relax}
\BIBdecl

\bibitem{assemblyline}
\BIBentryALTinterwordspacing
{0xADE1A1DE}, ``Assemblyline,'' 2022. [Online]. Available:
  \url{https://github.com/0xADE1A1DE/AssemblyLine}
\BIBentrySTDinterwordspacing

\bibitem{AhoSU86}
A.~V. Aho, R.~Sethi, and J.~D. Ullman, \emph{Compilers: Principles, Techniques,
  and Tools}, ser. Addison-Wesley series in computer science / World student
  series edition.\hskip 1em plus 0.5em minus 0.4em\relax Addison-Wesley, 1986.

\bibitem{variabilityCPU}
A.~R. Alameldeen and D.~A. Wood, ``Variability in architectural simulations of
  multi-threaded workloads,'' in \emph{{HPCA}}, 2003, pp. 7--18.

\bibitem{AlFardanP13}
N.~J. AlFardan and K.~G. Paterson, ``Lucky thirteen: Breaking the {TLS} and
  {DTLS} record protocols,'' in \emph{{IEEE} SP}, 2013, pp. 526--540.

\bibitem{sike}
\BIBentryALTinterwordspacing
R.~Azarderakhsh, M.~Campagna, C.~Costello, L.~D. Feo, B.~Hess, A.~Jalali,
  B.~Koziel, B.~LaMacchia, P.~Longa, M.~Naehrig, G.~Pereira, J.~Renes,
  V.~Soukharev, and D.~Urbanik, ``Supersingular isogeny key encapsulation --
  submission to the {NIST} post-quantum standardization project, round 2,''
  2019. [Online]. Available: \url{https://sike.org}
\BIBentrySTDinterwordspacing

\bibitem{Bansal06}
S.~Bansal and A.~Aiken, ``Automatic generation of peephole superoptimizers,''
  in \emph{{ASPLOS}}, 2006, pp. 394--403.

\bibitem{BERGMANN2003141}
S.~D. Bergmann, ``Compilers,'' in \emph{Encyclopedia of Information Systems},
  2003, pp. 141--170.

\bibitem{libsecp256k1}
\BIBentryALTinterwordspacing
{Bitcoin Core}, ``{libsecp256k1} - optimized {C} library for {ECDSA} signatures
  and secret/public key operations on curve secp256k1,'' 2022. [Online].
  Available: \url{https://github.com/bitcoin-core/secp256k1}
\BIBentrySTDinterwordspacing

\bibitem{Bokhari2020}
M.~A. Bokhari, B.~Alexander, and M.~Wagner, ``Towards rigorous validation of
  energy optimisation experiments,'' in \emph{{GECCO}}, 2020, pp. 1232--1240.

\bibitem{bokhari2019mind}
M.~A. Bokhari, L.~Weng, M.~Wagner, and B.~Alexander, ``Mind the gap - a
  distributed framework for enabling energy optimisation on modern smart-phones
  in the presence of noise, drift, and statistical insignificance,'' in
  \emph{IEEE CEC}, 2019, pp. 1330--1337.

\bibitem{CauligiSBJHJS17}
S.~Cauligi, G.~Soeller, F.~Brown, B.~Johannesmeyer, Y.~Huang, R.~Jhala, and
  D.~Stefan, ``{FaCT}: A flexible, constant-time programming language,'' in
  \emph{SecDev}, 2017, pp. 69--76.

\bibitem{COOPER2012597}
K.~D. Cooper and L.~Torczon, ``Chapter 11 - instruction selection,'' in
  \emph{Engineering a Compiler (Second Edition)}, 2012, pp. 597--638.

\bibitem{stabilizer}
C.~Curtsinger and E.~D. Berger, ``{STABILIZER:} statistically sound performance
  evaluation,'' in \emph{{ASPLOS}}, 2013, pp. 219--228.

\bibitem{doerr2019theory}
B.~Doerr and F.~Neumann, \emph{Theory of evolutionary computation: Recent
  developments in discrete optimization}.\hskip 1em plus 0.5em minus
  0.4em\relax Springer, 2019.

\bibitem{ErbsenPGSC19}
A.~Erbsen, J.~Philipoom, J.~Gross, R.~Sloan, and A.~Chlipala, ``Simple
  high-level code for cryptographic arithmetic - with proofs, without
  compromises,'' in \emph{{IEEE} SP}, 2019, pp. 1202--1219.

\bibitem{HARMAN2001833}
M.~Harman and B.~F. Jones, ``Software engineering using metaheuristic
  innovative algorithms: workshop report,'' \emph{Inf. Softw. Technol.},
  vol.~43, no.~14, pp. 905--907, 2001.

\bibitem{JoshiNR02}
R.~Joshi, G.~Nelson, and K.~H. Randall, ``Denali: {A} goal-directed
  superoptimizer,'' in \emph{{PLDI}}, 2002, pp. 304--314.

\bibitem{benchmarkPrecision}
T.~Kalibera, L.~Bulej, and P.~Tuma, ``Benchmark precision and random initial
  state,'' in \emph{SPECTS}, 2005, pp. 853--862.

\bibitem{KuepperJ23arXiv}
J.~Kuepper, A.~Erbsen, J.~Gross, O.~Conoly, C.~Sun, S.~Tian, D.~Wu,
  A.~Chlipala, C.~Chuengsatiansup, D.~Genkin, M.~Wagner, and Y.~Yarom,
  ``{CryptOpt}: Verified compilation with random program search for
  cryptographic primitives,'' ArXiv abs/2211.10665, 2022.

\bibitem{Massalin87}
H.~Massalin, ``Superoptimizer - {A} look at the smallest program,'' in
  \emph{{ASPLOS}}.\hskip 1em plus 0.5em minus 0.4em\relax {ACM}, 1987, pp.
  122--126.

\bibitem{MolnarPSW05}
D.~Molnar, M.~Piotrowski, D.~Schultz, and D.~A. Wagner, ``The program counter
  security model: Automatic detection and removal of control-flow side channel
  attacks,'' in \emph{{ICISC}}, 2005, pp. 156--168.

\bibitem{producingWrongData}
T.~Mytkowicz, A.~Diwan, M.~Hauswirth, and P.~F. Sweeney, ``Producing wrong data
  without doing anything obviously wrong!'' in \emph{{ASPLOS}}, 2009, pp.
  265--276.

\bibitem{PekhimenkoB10}
G.~Pekhimenko and A.~D. Brown, ``Efficient program compilation through machine
  learning techniques,'' in \emph{Software Automatic Tuning, From Concepts to
  State-of-the-Art Results}.\hskip 1em plus 0.5em minus 0.4em\relax Springer,
  2010, pp. 335--351.

\bibitem{threadContextSwitch}
K.~K. Pusukuri, R.~Gupta, and L.~N. Bhuyan, ``Thread tranquilizer: Dynamically
  reducing performance variation,'' \emph{{ACM} TACO}, vol.~8, no.~4, pp.
  46:1--46:21, 2012.

\bibitem{SasnauskasCCKTR17}
R.~Sasnauskas, Y.~Chen, P.~Collingbourne, J.~Ketema, J.~Taneja, and J.~Regehr,
  ``Souper: {A} synthesizing superoptimizer,'' Arxiv abs/1711.04422, 2017.

\bibitem{Schkufza0A13}
E.~Schkufza, R.~Sharma, and A.~Aiken, ``Stochastic superoptimization,'' in
  \emph{{ASPLOS}}, 2013, pp. 305--316.

\bibitem{Schkufza0A14}
E.~Schkufza, R.~Sharma, and A.~Aiken, ``Stochastic optimization of
  floating-point programs with tunable precision,'' in \emph{{PLDI}}, 2014, pp.
  53--64.

\bibitem{SharmaSCA13}
R.~Sharma, E.~Schkufza, B.~R. Churchill, and A.~Aiken, ``Data-driven
  equivalence checking,'' in \emph{{OOPSLA}}.\hskip 1em plus 0.5em minus
  0.4em\relax {ACM}, 2013, pp. 391--406.

\bibitem{SharmaSCA15}
R.~Sharma, E.~Schkufza, B.~R. Churchill, and A.~Aiken, ``Conditionally correct
  superoptimization,'' in \emph{{OOPSLA}}, 2015, pp. 147--162.

\end{thebibliography}

\end{document}